# Light Phase Detection with On-Chip Petahertz Electronic Networks

Yujia Yang[1], Marco Turchetti[1], Praful Vasireddy[1], William P. Putnam[1,2,3], Oliver Karnbach[1], Alberto Nardi[1], Franz X. Kärtner[1,3,4], Karl K. Berggren[1], Phillip D. Keathley[1]

[1]Research Laboratory of Electronics, Massachusetts Institute of Technology, Cambridge, MA, USA
[2]Department of Electrical and Computer Engineering, University of California, Davis, Davis, CA, USA
[3]Department of Physics and Center for Ultrafast Imaging, University of Hamburg, Hamburg, Germany
[4]Center for Free-Electron Laser Science and Deutsches Elektronen-Synchrotron (DESY), Hamburg, Germany

**Abstract**

**Ultrafast light-matter interactions lead to optical-field-driven photocurrents with an attosecond-level temporal response. These photocurrents can be used to detect the carrier-envelope-phase (CEP) of short optical pulses, and could be utilized to create optical-frequency, petahertz (PHz) electronics for information processing. Despite recent reports on optical-field-driven photocurrents in various nanoscale solid-state materials, little has been done in examining the large-scale integration of these devices. In this work, we demonstrate enhanced, on-chip CEP detection via optical-field-driven photocurrent in a monolithic array of electrically-connected plasmonic bow-tie nanoantennas that are contained within an area of hundreds of square microns. The technique is scalable and could potentially be used for shot-to-shot CEP tagging applications requiring orders of magnitude less pulse energy compared to alternative ionization-based techniques. Our results open new avenues for compact time-domain, on-chip CEP detection, and inform the development of integrated circuits for PHz electronics as well as integrated platforms for attosecond and strong-field science.**

## Introduction

In recent years, the combination of nano-optical structures with intense, few-cycle laser sources has led to a new class of solid-state petahertz electronic devices with promising applications in time-domain metrology as well as information processing[1–19]. These petahertz devices rely on the attosecond-level temporal response of optical-field-driven photocurrents that result from the interaction of strong electric fields (tens of GV/m) with nanostructured materials[2–6,8–10,13,15–17,19]. Unlike photocurrents in typical optoelectronic components, these optical-field-driven photocurrents are sensitive to changes in the electric field waveform of the optical pulse rather than the cycle-averaged photon density. Recent reports have demonstrated time-domain CEP detection with solid-state devices[6,13,15,16], and specifically, in Refs.[13,15] it was shown that photoelectron emission from plasmonic nanoantennas can be used to detect changes in the carrier-envelope phase (CEP).

Consider an optical pulse with a time-dependent electric field $F(t)=F_0(t)\cos(\omega t+\varphi_{ce})$ with a carrier angular frequency $\omega$, an intensity envelope $F_0(t)$, and a CEP $\varphi_{ce}$. The CEP determines the exact optical-field waveform of the pulse and has vital importance in ultrafast and strong-field nonlinear optical processes for few- to single-cycle pulses. In the time domain, the CEP is crucial for attosecond physics including ionization of atoms and molecules[20,21], high-harmonic generation[22], and attosecond pulse generation[23,24]. For frequency-comb sources, the carrier-envelope-offset (CEO) frequency, the frequency at which the CEP is oscillating, corresponds to a shift of the comb spectrum, and is important in applications such as optical frequency synthesis[25,26], high-precision metrology[27], and quantum information science[28].

Traditionally, the detection of the CEP has been achieved with both frequency-domain interferometric techniques[26,29,30] and time-domain photoelectron emission[31–36]. These traditional CEP measurement techniques either require frequency conversion and interferometry via multiple optical elements, or bulky vacuum apparatus and µJ-level pulses for the ionization of gas-phase atoms or molecules. A direct, time-domain CEP detection method using optical-field emission from nanoantennas[13,15] could enable shot-to-shot CEP tagging using orders of magnitude less pulse energy. Such a method holds promise for compact and on-chip CEP detectors operating in ambient conditions. However, due to: (1) low CEP-sensitivities, (2) material-damage thresholds, and (3) noise limitations; scaling the CEP-sensitive photocurrents to usable levels will require the synchronous operation of large-scale arrays of electrically-connected nanoantennas, in which case the photocurrent from individual nanoantenna adds up in phase at the read-out.

In this work, we fabricate and test large-scale networks of electrically-connected bow-tie nanoantenna pairs with nanoscale gaps for enhanced CEP detection (≈ 200 µm$^2$ array areas containing roughly 300 to 600 bow-ties). We demonstrate an order of magnitude improvement per emitter compared to single-triangle arrays with µm-scale emitter-collector spacing, and we show synchronous operation of the devices across the entire array, providing a route to shot-to-shot CEP tagging of nJ-level pulses. We address key challenges in both the design and the operation of such large-scale, electrically-connected arrays including electromagnetic sensitivity to design parameters, *in-situ* removal of electrical shorts caused by process variations, and noise sources that limit the devices' ultimate signal-to-noise ratio. We conclude that we are operating the devices at or near their ultimate noise floor set by the shot-noise arising from the total number of emitted electrons. Beyond CEP detection, this work has ramifications for the development and understanding of electrically integrated nanoantenna devices for on-chip attosecond science and PHz electronics.

**Results**

Fig. 1a shows a schematic of the nanoantenna device used in this work. An array of plasmonic bow-tie nanoantennas was supported by a transparent, insulating substrate. Each bow-tie nanoantenna consisted of a pair of nanotriangles. All of the left nanotriangles of the bow-ties were electrically connected to one contact pad, while all right nanotriangles were electrically connected to another contact pad. The device was essentially a parallelized array of photoelectron tunneling devices[13]. For a bow-tie nanoantenna, the two nanotriangles were the cathode and anode, for photoelectron emission and collection respectively; and in our devices, the cathode-to-anode gap was in the range of 10-50 nm. These devices operated in ambient conditions thanks to the nanoscale cathode-to-anode gap.

Our configuration illustrated in Fig. 1a has several important advantages. First, the nanometer-scale gap ensured sub- to few-femtosecond transit times of electrons between emitters. This rapid transit time enabled hundreds of THz- to PHz-level operating bandwidths, reduced the electron's interaction with gas molecules in the ambient environment, and removed the need for large bias voltages to collect the emitted electrons. Second, by directing the photocurrent with integrated connecting wires, the signal could either be accumulated or selectively coupled to down-stream electronics on a femtosecond timescale[17]. Third, the inversion symmetry of the bow-tie devices resulted in a balanced detection scheme whereby the CEP-sensitive signal was retained while the total average current was canceled, reducing background noise due to laser intensity fluctuations and enabling detection with high-gain amplifiers. Finally, the connecting wires and nanoantennas could all be produced with a single lithography step, simplifying fabrication and ensuring nanometer-level alignment accuracy between the emitters and connecting wires, which we will show to be critical to device operation.

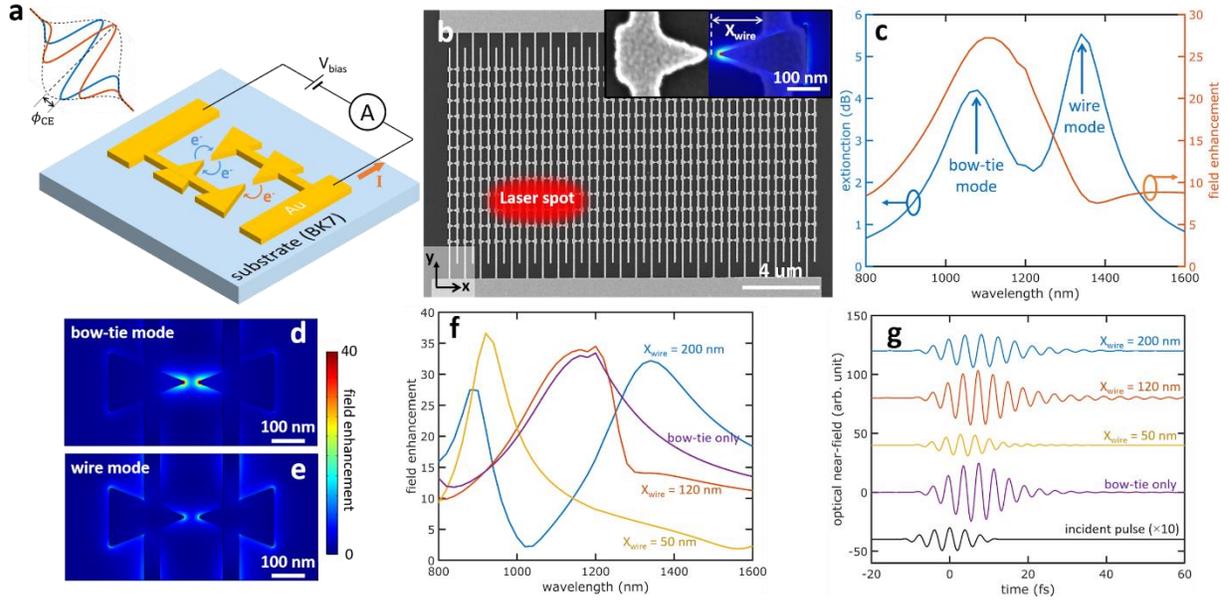

**Figure 1**. Electrically connected plasmonic bow-tie nanoantenna arrays. **a**, Schematic diagram of the gold bow-tie nanoantenna devices on insulating substrate ($V_{bias}$, bias voltage; A, ampere meter; $I$, net photocurrent). Left nanotriangles are electrically connected to one contact pad, while right nanotriangles are electrically connected to the other. An incident ultrafast optical pulse induces photoelectron emission across the nano-gaps between the nanotriangles. The carrier-envelope-phase (CEP) $\varphi_{ce}$ of the pulse affects the photocurrent measured in the external circuit. For $\varphi_{ce}=\pi/2$ (blue trace), the pulse has two symmetric optical half cycles and the photocurrent in the two opposite directions cancel each other, leading to a zero net photocurrent. For $\varphi_{ce}=\pi$ (orange trace), the pulse has only one strong optical half cycle, and a net photocurrent can be measured. In the experiment, $\varphi_{ce}$ is modulated by an oscillating signal with a carrier-envelope-offset (CEO) frequency $f_{ceo}$. This CEO frequency can be measured from the photocurrent spectrum. **b**, SEM image of a plasmonic nanoantenna array consisting of 288 bow-tie nanoantennas. The laser beam spot size in the experiment is also illustrated. The nanoantenna arrays are on silicon substrates for imaging the devices without charging issues. Inset: SEM image of a plasmonic bow-tie nanoantenna with a nano-gap of 28 nm. The superimposed color plot shows the simulated optical near-field profile (showing a spatial map of the electric field magnitude normalized by the incident electric field magnitude) of a nanoantenna with similar dimensions. **c**, Simulated extinction spectrum (blue) and field-enhancement spectrum (orange) of the electrically connected bow-tie nanoantenna array. The two extinction peaks are labeled as the bow-tie mode and the wire mode. **d**&**e**, Simulated optical near-field profiles (showing spatial maps of the electric field magnitude normalized by the incident electric field magnitude) of the bow-tie mode and the wire mode. The color scale is saturated for better visualization. **f**, Simulated field-enhancement spectra of the plasmonic bow-tie nanoantenna arrays with different connecting wire positions ($X_{wire}$ labeled in **b**, representing the x-distance between the inner edge of the wire and the center of the bow-tie structure). For comparison, the field-enhancement spectrum for a bow-tie nanoantenna array without the connecting wires is also shown. **g**, Simulated time-domain response of the plasmonic bow-tie nanoantenna arrays with different connecting wire positions. The waveforms show the optical field at the nanotriangle tip.

The nanoantennas enhance the electric field of ultrafast optical pulses, inducing photoelectron emission. With a light polarization along the bow-tie axis of symmetry in the x-direction, the optical field induces photoemission current flowing between the two nanotriangles within a bow-tie nanoantenna. The emitted electrons transit from the nanotriangles on one side to the nanotriangles on the other side, and vice versa as the optical field switches its direction. The photocurrent is then measured by connecting the two contact pads to an external ammeter and voltage bias. The CEP $\varphi_{ce}$ of the optical waveform controls the amplitude and direction of the induced photocurrent. For example, given a waveform represented by the blue trace in Fig. 1a, when $\varphi_{ce}=\pi/2$, the pulse has two symmetric optical half cycles, and the photocurrent in the two opposite directions cancel each other, leading to zero net photocurrent. When $\varphi_{ce}=\pi$ (represented by the orange trace in Fig. 1a), the pulse has only one strong optical half cycle (the contribution from the weak optical half cycles is negligible due to the nonlinearity of the photoemission process), and a net photocurrent is generated which flows from the left side of the excited bow-ties to the right side. Likewise when $\varphi_{ce}=2\pi$, the same photocurrent is generated only now flowing from the right side to the left.

Fig. 1b shows the SEM image of a nanoantenna array consisting of 24×12 bow-ties, with the full array covering an area of about 20×10 μm². The nanoantenna array is fabricated on a silicon substrate (for imaging purpose), which is conductive and free from charging issues, under the same conditions for nanoantenna fabrication with glass substrates. The inset of Fig. 1b shows a plasmonic bow-tie nanoantenna with a nano-gap of 28 nm (see Supplementary Note 1 for nano-gap size measurement). The superimposed color plot shows the simulated optical near-field profile of a nanoantenna with similar dimensions, featuring an enhanced optical field at the nano-gap.

Fig. 1c-e show the simulated optical response of a nanoantenna array. Fig. 1c shows the extinction and field-enhancement spectra. The dimensions of the nanoantenna array were taken from the SEM images of a fabricated sample (see Supplementary Note 2). The extinction is defined as $-10log_{10}(T/T_0)$, where $T$ is the power transmissivity when the nanoantenna is present, and $T_0$ is the power transmissivity when the nanoantenna is absent (but the substrate is still present). The field-enhancement is defined as the ratio of the optical near-field at the nanotriangle tip near the nano-gap (referred to as the "tip" in the following), averaged over the curved surface defined by the tip radius of curvature and gold thickness, to the optical field of the incident light. It can be seen that the extinction spectrum shows a double-peak feature with one peak at a wavelength of ≈ 1100 nm and another peak at a wavelength of ≈ 1300 nm. The optical near-field around the nanoantenna is plotted at 1100 nm (Fig. 1d) and 1300 nm (Fig. 1e) incident wavelengths. For the peak around 1100 nm, the optical field is localized at the bow-tie tip, indicating this peak shows the plasmonic resonant mode of the bow-tie nanoantenna, and we named this mode the "bow-tie mode". For the peak around 1300 nm, the optical field is enhanced both around the bow-tie nanoantenna as well as the connecting wires, indicating a plasmon mode propagating and resonating along the periodic array of bow-ties and wires; we refer to this mode as the "wire mode". We emphasize that the wire mode is not localized and travels along the wires throughout the periodic lattice of devices, and as such depends on the periodic nature of the device layout. The wire mode has a relatively weak field enhancement at the nanotriangle tip. On the other hand, the bow-tie mode is localized to the bow-tie antenna and contributes strongly to the field-enhancement at the nanotriangle tips (see the field-enhancement

spectrum in Fig. 1c and the optical near-field profiles in Fig. 1d&e). As a result, the bow-tie mode has a much stronger influence on the photoelectron emission compared to the wire mode.

To investigate the effect of the electrical connecting wires on the nanoantenna optical response, we performed simulations with different connecting wire positions (Fig. 1f&g). Fig. 1f shows the simulated field-enhancement spectra of the plasmonic bow-tie nanoantenna arrays with different connecting wire positions ($X_{\text{wire}}$ labeled in Fig. 1b). For comparison, we also include the spectra for the bow-tie nanoantenna without connecting wires. The bow-tie nanoantenna without wires shows a single-peak field-enhancement and extinction spectra (the extinction spectra are shown in Supplementary Fig. 2). When the connecting wires are added, the field-enhancement spectra splits into two peaks, corresponding to the aforementioned bow-tie and wire modes. The spectral separation of the two peaks is small and they merge into a single peak, if the connecting wire position is near the center of the nanotriangle (e.g. $X_{\text{wire}}$ = 120 nm). The spectral separation of the two peaks increases, with the bow-tie mode being blue-shifted and the wire mode being red-shifted, when the connecting wire position is close to the nanotriangle base (e.g. $X_{\text{wire}}$ = 200 nm) or nanotriangle tip (e.g. $X_{\text{wire}}$ = 50 nm; the second peak is beyond the displayed spectral range). For the extinction spectra (Supplementary Fig. 2), a similar behavior is observed. In general, the bow-tie plasmonic mode is least disturbed when the connecting wire is close to the nanotriangle center where there is a node of the optical near-field distribution[37,38]. Fig. 1g shows the simulated time-domain response at the tips of the plasmonic bow-tie nanoantennas with different connecting wire positions assuming a $\cos^2$-shaped incident pulse with a central wavelength of 1177 nm and a pulse duration (FWHM) of 10 fs (analogous to the experimental pulses we use in this work). While the broadband plasmonic enhancement preserves the ultrafast character of the incident pulse, the wire position clearly influences the time-domain profile of the waveform, and thus the photoemission response. Hence, it is critical that the wire position is uniform to ensure device response uniformity. As with the frequency-domain response, the highest field-enhancement was obtained by placing the connecting wires near the center of the nanotriangles.

The fabricated bow-ties had a distribution of gap sizes that differ from the nominal size due to process variations. For nominal gap sizes on the order of 10 nm or less, it is not uncommon for several antennas in a column to be connected together (*i.e.* no gap), with the cathode electrically shorted to the anode. This shorting makes it impossible to measure any generated photocurrent. To resolve this issue we used electromigration to break the connecting wires of these shorted columns, thus removing them from the circuit (see Fig. 2 and Supplementary Note 3). A bias voltage was applied to the nanoantenna array creating a large current density in the wires of the shorted columns. By applying sufficient voltage to the array, and thus generating sufficiently high current densities, the connecting wires of these shorted columns were broken. However, columns having no shorted devices exhibited a very high effective resistance and were left unchanged. Fig. 2a shows the voltage and current across a nanoantenna array during electromigration. The average current to initiate electromigration was ~1 mA per connecting wire, with a current density (~1 A/µm$^2$) consistent with previous reports on current-induced electromigration in gold nanowires[39,40]. Fig. 2b shows the SEM image of a shorted nanoantenna array after electromigration where the connecting wires have been broken. Fig. 2c&d show non-shorted and shorted columns of bow-tie nanoantennas respectively from another array. If all bow-ties along the column were disconnected (Fig. 2c), the column was an open-circuit and the connecting wires were not broken. However, if there were connected bow-ties along the column (Fig. 2d), the column was shorted and the connecting wires were broken by electromigration. Hence, we were able to use this electromigration process as a surgical tool to selectively remove shorted devices from an array when needed.

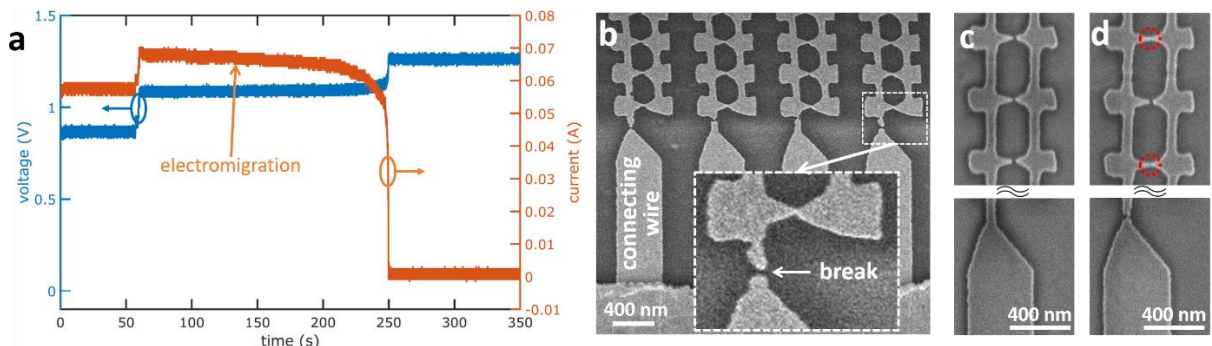

**Figure 2.** Electromigration of electrically connected nanoantenna arrays. **a**, Applied voltage (blue) and current (orange) across a plasmonic nanoantenna array during the electromigration process. Electromigration transformed a short-circuit array into an open-circuit array. **b**, SEM image of a connected plasmonic nanoantenna array after electromigration. The electrical connecting wires were broken and disconnected during electromigration. Inset: zoomed-in image of the connected bow-tie nanoantenna and the broken connecting wire. **c**, If all the bow-tie nanoantennas along a connecting wire were disconnected to begin with, the wire was not affected and remained intact. **d**, If there were shorted bow-tie nanoantennas (red dashed circles) along a connecting wire, the wire was broken via electromigration, eliminating the corresponding nanoantenna column from the functional device. In **c**&**d**, only part of the bow-tie nanoantennas along a connecting wire are shown in the image (but all the nanoantennas were inspected with the SEM).

To test the CEP-sensitivity of the devices, we used a CEP-stabilized supercontinuum source similar to that described in prior work[15,41] and briefly described in the Methods. The supercontinuum source has a central wavelength of ≈1177 nm, pulse duration of ≈ 10 fs full width at half maximum (FWHM) (≈ 2.5 cycles FWHM), repetition rate of 78 MHz, and peak pulse energy of ≈ 190 pJ. The carrier-envelope-offset frequency $f_{ceo}$ was stabilized to 100 Hz, meaning that $\varphi_{ce}$ of each pulse was shifted by a constant amount such that $\varphi_{ce}$ of the $n$-th pulse was $\varphi_{ce}[n]=2\pi n f_{ceo}/f_{rep}+\varphi_0$, where $n$ is the pulse number and $\varphi_0$ is a constant phase offset. The beam was focused to a spot size of 2.25 μm ×4.1 μm FWHM resulting in a peak intensity of ≈ $2.6 \times 10^{11}$ W/cm$^2$ (corresponding to a peak field of ≈ 1.4 GV/m) before enhancement. To characterize the CEP-sensitive current response of the device array, the amplitude and phase of the current response at 100 Hz was measured via lock-in detection (see Supplementary Note 4). During the measurement, a barium fluoride (BaF$_2$) wedge was translated through the beam every 20 seconds providing discrete shifts (measured as ∼54.4° ± 11°, calculated as ∼57.9°) in the CEP allowing us to verify that the measured current response was indeed CEP-sensitive. The optical power absorption in the wedge is negligible. The measurement results of this scan are shown in Fig. 3.

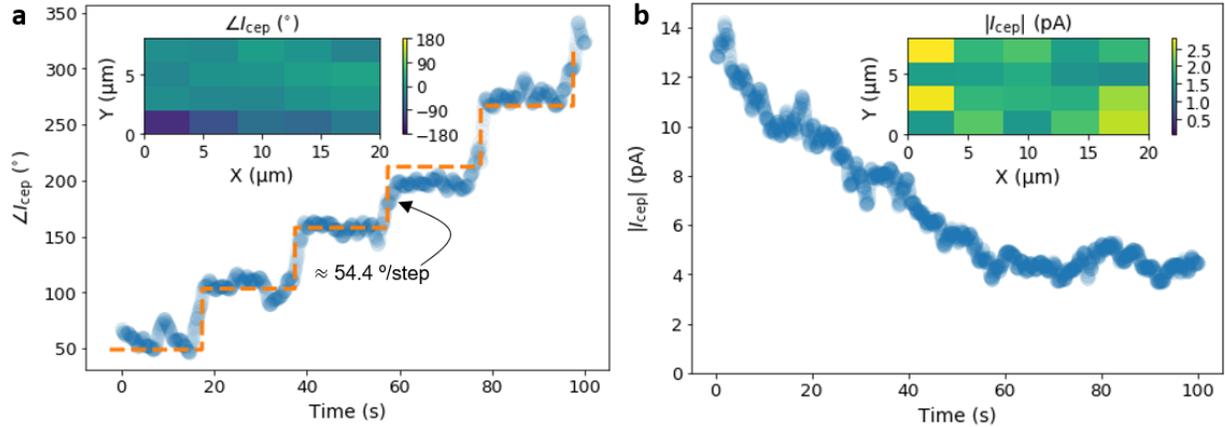

**Figure 3**. Carrier-envelope-phase-sensitive current from electrically-connected bow-tie arrays. **a**, Phase of $I_{cep}$ as a function of time. A barium difluoride (BaF$_2$) wedge is inserted into the beam every twenty seconds leading to a measured average shift in $\varphi_{ce} \approx 54.4° \pm 11°$. The orange dashed trace shows a staircase fit to the measured data. The inset shows the average phase value while scanning over the entire array area. **b**, Corresponding value of $|I_{cep}|$ over the same scan shown in **a**. The optical power absorption in the wedge is negligible. The inset shows the amplitude of $I_{cep}$ while scanning the beam over the entire array.

The peak CEP-sensitive current measured was $\approx$ 14 pA at 100 Hz for an array with $\approx$ 1.5625 nanoantennas/μm², which corresponds to 1.12 electrons/pulse. Considering that $\approx$ 11 bow-tie pairs were exposed within the FWHM of the beam spot, this corresponds to roughly 1.3 pA/bow-tie, which is similar to the results in Ref.[13] (roughly 0.6 pA from a single bow-tie nanoantenna), and constitutes more than one order of magnitude increase in CEP-sensitive current compared to similar single-nanotriangle emitters we have reported on in prior work (up to 1.5 pA CEP-sensitive current)[15,19]. Despite these encouraging results, we observed a rather fast degradation in this current response over a period of tens of seconds before eventual stabilization to a current level of $\approx$ 4 pA ($\approx$ 0.36 pA/bow-tie). Nevertheless, due to the combined scalability of the array configuration and benefits of the nanoscale emitter-collector separation of our devices, both the peak and the stabilized optical-field-sensitive photocurrents represent a significant improvement compared to the current generated by a single bow-tie nanoantenna[13] or plasmonic nanoparticles with mesoscopic emitter-collector gaps[15].

To demonstrate reliable operation across the entire emitter array, and the potential for signal multiplexing by interconnecting devices, we rastered the beam spot across the connected nanoantenna array while collecting the CEP-sensitive current amplitude $|I_{cep}|$ and phase $\angle I_{cep}$. These results are shown in the insets of Fig. 3. Despite active and inactive spots in the array due to the nonlinear dependence of the photoemission on the emitters' precise shape and surface properties, the scan shows that the entire array was active, with an average CEP-sensitive current response of 1.5 ± 0.8 pA. More importantly, the phase variation of the CEP-sensitive response across the entire array was only ±42°(±733 mrad) (the inset of Fig. 3a), indicating that by illuminating larger areas of the array while holding the peak intensity fixed (see Supplementary Note 5), one could scale the CEP-sensitive current by area with 80% efficiency (*i.e.* using a beam spot of *X* greater area would result in a CEP-sensitive current increase of 0.8$X$). However, we point out that in the inset of Fig. 3a, the lowest row of data points differs significantly from the upper rows. This difference could be because our laser spot was close to the edge of the array. Excluding this row, we find that the phase variation reduces to ±26° (±454 mrad), with a scaling efficiency of 90%.

With a similar illumination intensity and a low phase variation, the nanoantennas connected in parallel within an array can achieve synchronous operation. We calculate that with pulse energies on the order of 10-100 nJ spread across similarly sized arrays one could achieve peak current signals of sufficient level for single-shot CEP-tagging (see Supplementary Note 6; note the calculation is based upon stabilized, rather than peak, CEP-sensitive signal to ensure stabilized operation of the device over a long time). Furthermore, we should also note that the CEP-sensitive current was measured for several samples from multiple fabrication batches, and the effects of nano-gap size and laser pulse energy were investigated (see Supplementary Note 8). The degradation of CEP-sensitive signal was observed for several devices, especially those with a small gap size illuminated with a high laser pulse energy. However, once stabilized at a lower signal level, the devices operated for hours without further degradation.

We routinely observed the aforementioned photocurrent degradation for small nano-gap devices under a high pulse energy (the degradation was observed for pulse energies in the range of 140-190 pJ; the exact pulse energy differs for different devices). Fig. 4a&b show the SEM images of a nanoantenna array before (Fig. 4a) and after (Fig. 4b) high-intensity illumination for photoemission measurements. The average gap size of the bow-tie nanoantennas increased from 50.5±3.2 nm to 62.2±11.9 nm (see Supplementary Note 8 for measurements on more samples). To investigate the impact of this nanoantenna reshaping, we performed optical extinction measurements on nanoantenna arrays both before and after high-intensity illumination.

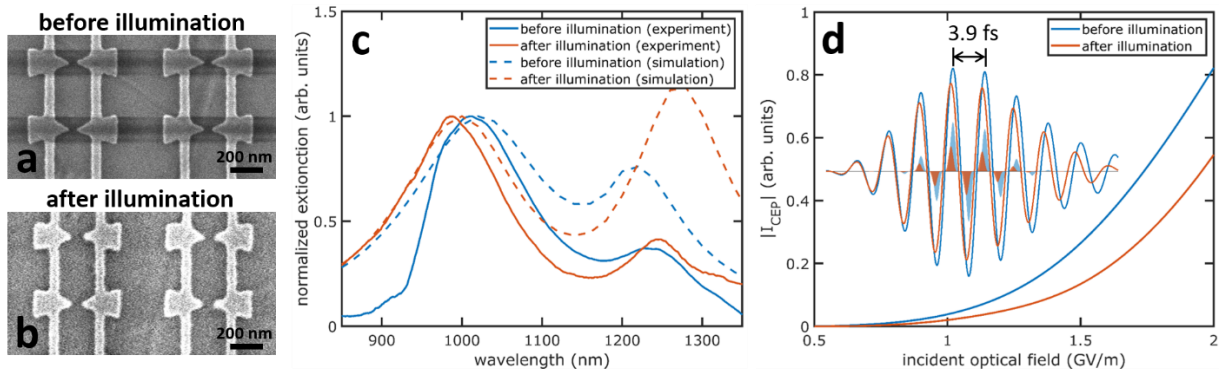

**Figure 4**. Nanoantenna device degradation during photoemission measurement. **a&b**, SEM images of a bow-tie nanoantenna array before and after photoemission measurement. The average gap size of the bow-tie nanoantennas increased from 50.5±3.2 nm to 62.2±11.9 nm after illumination. The contrast variation is caused by charging issues of the insulating substrate during imaging. **c**, Measured and simulated extinction spectra of the nanoantenna array shown in **a&b** before and after photoemission measurement. **d**, Simulated CEP-sensitive photocurrent magnitude |$I_{cep}$| *versus* the optical near-field for the nanoantenna array before and after photoemission measurement. Inset:

Simulated time-domain response of the nanoantenna array before and after photoemission measurement. The waveforms show the plasmonically enhanced optical fields at the nanotriangle tip for the nanoantenna arrays before and after illumination. The shaded areas show the waveforms of the photoemission current calculated from the Fowler-Nordheim theory.

Fig. 4c shows the simulated and measured extinction spectra from an array before and after photoemission measurements were performed. The simulation used geometries extracted from SEM images of the array (Fig. 4a&b). The double-peak feature of the spectra was obtained in both simulation and measurement, with the bow-tie mode around 1050 nm wavelength and the wire mode around 1250 nm wavelength. We note that the wire mode was significantly stronger in simulation than in measurement. We attribute this to the fact that the simulation used periodic boundary conditions and assumed the wires were infinitely long and the arrays perfectly periodic, while the fabricated devices consisted of finite arrays and wires with imperfect periodicity. The bow-tie mode on the other hand is associated with a localized surface plasmon resonance of the individual nanoantennas, and was thus better represented in simulation. Several other features of the extinction spectra were reproduced in simulation and measurement. For instance, the exposure to optical pulses led to a larger spectral separation between the two peaks: the bow-tie mode was slightly blueshifted, while the wire mode was slightly redshifted, with an increase in its intensity. The change of the extinction spectra before and after photoemission measurement is caused by a combined effect of laser-induced reshaping of the bow-ties and the change of the relative position between the wires and the nanotriangles (see earlier discussions) as a result of the reshaping.

To simulate the effect of the reshaping on the CEP-sensitive photoemission, the photocurrent was estimated using a quasi-static Fowler-Nordheim tunneling theory[19]. Fig. 4d shows the simulated CEP-sensitive photocurrent magnitude $|I_{cep}|$ with a varying peak incident optical-field strength for the nanoantenna array before (blue) and after (orange) photoemission measurements. The inset shows the plasmonically-enhanced optical-field waveforms (solid curves) and the calculated time-dependent photoemission current density (shaded area). This calculation confirms that a decreased field-enhancement, caused by laser-induced reshaping of the plasmonic nanostructures, is the dominant factor causing the reduced CEP-sensitive photocurrent. The result is in qualitative agreement with experimental observations. However, it is not a general result that the CEP-sensitivity always reduces as a result of the emitter reshaping as there is a complicated interplay between the CEP-sensitivity and the emitter resonance and peak intensity (see Supplementary Note 9 and Ref.[19]).

For reliable operation of such nanoantennas as CEP detectors or optical-field-driven circuit elements, it is critical to understand their noise characteristics and how the signal-to-noise ratio (SNR) could be further enhanced for subsequent amplification and processing. A typical photocurrent frequency spectrum near $f_{ceo}$ Hz is shown in Fig. 5a, exhibiting $|I_{cep}| \approx$ 4 pA, and a SNR of $\approx$ 254 ($\approx$ 25 dB at 0.5 Hz resolution bandwidth). The balanced devices used in this work should ideally reduce sensitivity to pulse energy fluctuations, as the total average current response of each nanotriangle emitter in the bow-tie pair cancels. To ensure this is the case, we performed the photocurrent measurement with a varying DC bias voltage between the two nanotriangle emitters in the bow-tie pair. Specifically, we measured the photocurrent response at 0 Hz corresponding to the total average current collected $I_{0,\text{collected}}$ as a function of DC bias voltage $V_{\text{bias}}$. In unbalanced configurations, $I_{0,\text{collected}}$ contains a significant portion of photocurrent that depends only on the optical pulse intensity and is not sensitive to the CEP[19]. As shown in Fig. 5b, the bias voltage can indeed control the amount of average DC current collected from the devices by breaking the symmetry between the bow-tie pairs. Importantly, there is a point where $I_{0,\text{collected}}$ is nearly eliminated. As one might expect, the $V_{\text{bias}}$ value that gives $I_{0,\text{collected}} \approx 0$A varies slightly across the sample (sometimes slightly positive, sometimes slightly negative; see Supplementary Note 10) due to natural asymmetries from the fabrication process. Nonetheless, it was found that the noise level was insensitive to this $V_{\text{bias}}$ setting, and that the SNR of the devices tested here was strikingly similar to prior measurements performed with asymmetric single nanotriangle nanoemitters despite a reduction of $I_{0,\text{collected}}$ by more than two orders of magnitude for the case of the symmetric bow-tie pairs (see Supplementary Note 10). This similarity of the SNR indicates that the noise floor measured in both symmetric and asymmetric cases was not a result of common-mode noise in the $I_{0,\text{collected}}$ signal, *e.g.* noise from fluctuations of the incident optical pulse energy, as the symmetric bow-tie configuration formed a balanced detection scheme that should significantly reduce the common-mode noise (see further discussion and comparison to asymmetric triangular devices in Supplementary Note 10).

To investigate the behavior of the noise-floor under illumination, we characterized the root-mean-square (RMS) average noise current $I_{\text{noise}}$ as a function of both the pulse energy (and thus the peak intensity) and $I_{0,\text{collected}}$ while setting $V_{\text{bias}}$=3 V. (Note that we chose to operate the devices under a bias voltage such that $I_{0,\text{collected}} \neq 0$ so that it could be monitored). For these measurements $f_{\text{ceo}}$ was unlocked, and $I_{0,\text{collected}}$ was measured by chopping the beam and measuring the current at the chopping frequency, which was set between 100-150 Hz. When examining $I_{\text{noise}}$ as a function of $I_{0,\text{collected}}$ at various frequency locations using multiple device arrays, a square-root dependence was consistently observed. Given the square-root dependence, in Fig. 5c we plot $I_{\text{noise}}$ vs. $I_{\text{eq}} = \alpha I_{0,\text{collected}}$ for two separate device arrays (referred to as Array 1 and Array 2). We define $I_{\text{eq}}$ as the equivalent shot-noise current source such that $I_{noise} = \sqrt{2q\Delta f I_{eq}}$, with $q$ being the electron charge and $\Delta f$ being the resolution bandwidth of the measurement. For Array 2 we analyze $I_{\text{noise}}$ at 100 Hz and 340 Hz for comparison. To determine α, we fit the measured data in each case to the reference line set by $\sqrt{2q\Delta f I_{eq}}$, which is shown in orange.

The noise amplitude measured was too strong to be accounted for by the weak values of $I_{\text{cep}}$ observed, and was found to be uncorrelated to the strength of the CEP note (see Supplementary Note 11). Thermal noise was ruled out as the scaling of the noise with incident pulse energy was relatively uncorrelated across the tested arrays, and did not scale at the expected rate relative to the incident pulse energy (see Supplementary Note 11 for further data and explanation). In Fig. 5d we examine only the noise power spectral density as a function of frequency (corresponding to Array 2 from Fig. 5c). We note that until around 200 Hz the noise scales as $1/f^{3/2}$. This noise scaling has been observed in field-emission devices and is often attributed to work-function fluctuations resulting from Brownian motion of impurities on the metal surface[42–45]. After 200 Hz there is a transition to a spectrally flat noise response that is still well above the noise floor of the detector (Supplementary Fig. 10). We note that this transition to the spectrally flat region can shift from sample to sample and spot to spot (see for instance Supplementary Fig. 10 where this transition is closer to 150 Hz) but appears to be a general behavior of the noise spectrum from our devices. This spectrally flat noise response means that simply shifting to higher values of $f_{ceo}$ would not significantly improve the SNR of the devices, and that, due to the observed square-root dependence of the noise on the collected current, such electrically connected networks and larger beam spots for increased signal are critical to improving the SNR.

We attribute the flat spectral region in Fig. 5d to the shot-noise floor of the devices arising from the total *emitted* current $I_{0,emitted}$ (that is the sum of all charge *emitted* from every nanotriangle emitter on the sample surface, including the charge that is not detected due to the cancellation of the net photocurrent by balanced detection). Since the shot-noise of electron emission from independent nanotriangle emitters in bow-tie pairs would not be correlated, and would thus not cancel despite the inversion-symmetric bow-tie arrangement, the RMS average shot noise current should scale with the total *emitted* current such that $I_{eq}=I_{0,emitted}$ in this region, *i.e.* $I_{noise} = \sqrt{2qI_{0,emitted}\Delta f}$ (of course, as we discuss in Supplementary Note 7, this may not hold for cases where the emission from both sides would be correlated, such as from a single quantum electron state extending across the gap). The value of $I_{0,emitted}$ and the precise CEP sensitivity can vary from device to device, and might explain the shift in the transition frequency from $1/f^{3/2}$ scaling to shot-noise from spot to spot.

To support our interpretation that shot-noise from $I_{0,emitted}$ is the cause of the observed flat noise-floor beyond 200 Hz, we used the data from 340 Hz on Array 2 shown in Fig. 5c to estimate that $I_{0,emitted} \approx 40$ nA at the highest pulse energy tested. Given the strength of the CEP-sensitive photocurrent for this array at the peak tested pulse energy, we calculate a CEP-sensitivity $I_{cep}/I_{0,emitted}$ between $10^{-4}$ to $10^{-5}$. Both the estimated $I_{0,emitted}$ and CEP-sensitivity values are in good agreement with prior results using single nanotriangle emitters made of the same material with similar tip geometries, peak enhancement factors and the same laser source, lending extra confidence to the conclusion that the SNR of the CEP response of devices was measured either just above or at the shot-noise-limit[15,19]. This shot-noise-limit emphasizes the importance of using large scale device arrays, both for improving the signal, and the SNR, which should increase with $\sqrt{I_{0,emitted}}$. Considering the illumination of an entire array having dimensions of 50×50 µm², we find that it would be possible to achieve an SNR of 20-30 dB at a resolution bandwidth of 1 kHz (see Supplementary Note 6). With further improvement of the detector area or the nanoantenna density, the improved SNR could be sufficient for feedback and control of $f_{ceo}$.

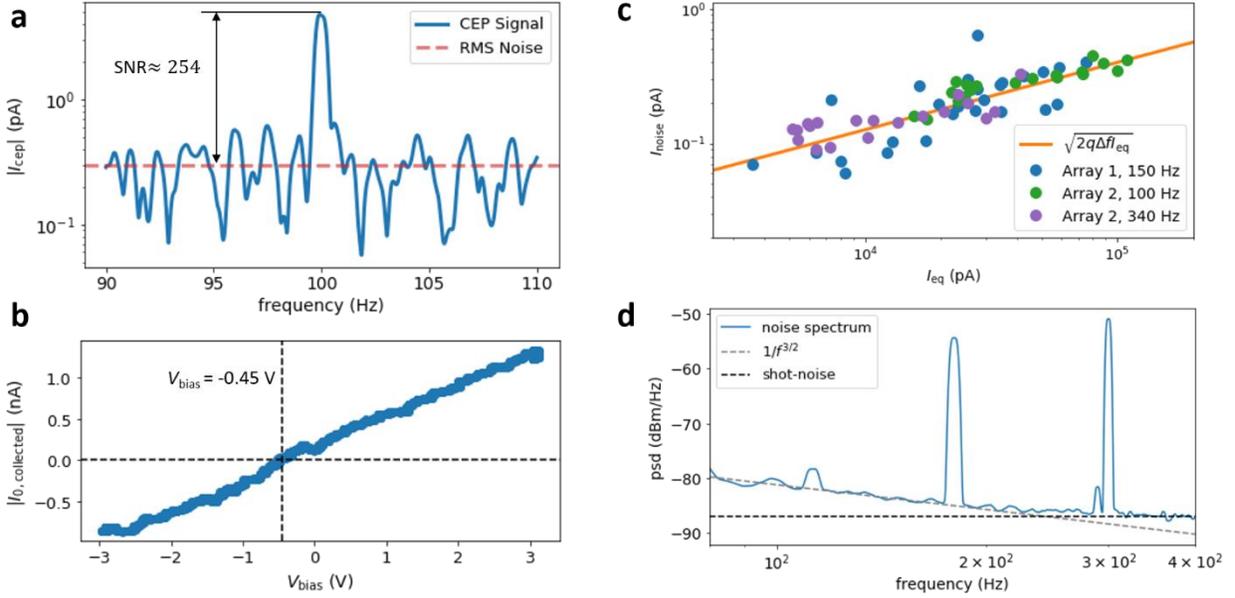

**Figure 5**. Results demonstrating SNR, balanced configuration, and noise characterization. **a,** Spectrum of the photocurrent near $f_{ceo}$ = 100 Hz, showing the CEP-sensitive current spike and surrounding noise with an SNR of ≈ 254 at 0.5 Hz resolution bandwidth. **b,** Plot of $I_{0,\text{collected}}$ demonstrating the use of $V_{bias}$ to null the total average collected current signal. The signal is nulled when $V_{bias}$ ≈ -0.45 V. **c,** Plot of the noise signal $I_{noise}$ vs. the equivalent shot noise current $I_{eq}$ for two arrays while setting $V_{bias}$ = 3 V. Blue dots represent results from Array 1 at 150 Hz, the green dots represent Array 2 at 100 Hz, and the purple dots Array 2 at 340 Hz. The orange reference curve represents $\sqrt{2q\Delta f I_{eq}}$. **d,** Power spectral density (psd) as a function of frequency for Array 2 of **c**. The CEP was unlocked. Reference curves demarcate $1/f^{3/2}$ scaling and the shot-noise floor respectively. The visible notes are due to background power-line noise (at 120 Hz, 180 Hz and 300 Hz respectively).

## Discussion

In this work, we have demonstrated on-chip medium-to-large-scale integration of nanoscale optical-field-driven devices. We have investigated device design, fabrication, multiplexing, degradation, and noise characteristics. In the current design, nearly identical individual devices are electrically connected in parallel and operated synchronously to produce a large optical-field-sensitive photocurrent. Noise analysis shows that we are operating these devices near their shot-noise limit which arises due to the average total emitted current signal from each emitter in the bow-tie pairs despite the fact that there is no net collected average total signal. Our findings emphasize the need for large-scale arrays such as those investigated here to further improve the overall SNR and achieve sufficient photocurrents for shot-to-shot CEP tagging. To that end, we find that by illuminating arrays of similar dimensions to those currently fabricated with pulses having an energy of just 10-100 nJ and similar duration to that used in our experiments, one could achieve sufficient photocurrent for shot-to-shot CEP tagging and an improvement in SNR of two to three orders of magnitude. Compared to current state-of-the-art field-ionization CEP detectors[32], such devices reduce the needed pulse energy by at least two to three orders of magnitude (see Supplementary Note 6), while replacing bulky vacuum equipment and electron detectors with a simple, monolithic optoelectronic device operating in ambient conditions. These performance improvements could have a significant impact on experiments and applications requiring optical waveform control and synthesis. In order to operate the CEP detectors at even higher power for potentially greater signal yield and SNR without device degradation, alternative refractory plasmonic materials[46,47] could be considered. Additionally, the nanoantenna arrays demonstrated in this work can be viewed as optical-field-driven, PHz integrated circuits with identical individual devices. By modifying the interconnection and integrating heterogeneous devices and materials, this study will inform the development of more complex integrated circuits of PHz electronics[48-50], as well as on-chip integrated platforms for attosecond and strong-field science.

**Methods**

**Device fabrication**
The plasmonic bow-tie nanoantenna arrays were fabricated on glass (BK7) substrates (*MTI Corp.*) with electron beam lithography (EBL) and a metal liftoff process. A ~70 nm film of poly(methyl methacrylate) (PMMA) (*MicroChem Corp.*) was spin-coated onto the substrate and then soft-baked at 180 °C. A thin layer of Espacer (*Showa Denko*) was then spin-coated for charge dissipation during electron beam lithography. Bow-tie nanoantenna and electrical connecting wire patterns were produced by an Elionix F-125 electron beam lithography system using an accelerating voltage of 125 kV and a beam current of 500 pA. The bow-tie nanostructures and electrical connecting wires were defined and fabricated in one EBL step instead of two aligned EBL steps. Fabrication of the two structures together ensured good alignment accuracy between the bow-ties and the connecting wires, which is critical for tuning the optical response of the nanoantenna arrays as shown in the main text. After exposure, Espacer was removed with 60 s DI water rinse. Exposed PMMA was developed in 3:1 isopropyl alcohol (IPA) : methyl isobutyl ketone (MIBK) at 0 °C for 30 s and then dried with flowing nitrogen gas. 2 nm Ti and 20 nm Au were then deposited via electron-beam evaporation. Metal lift-off was performed in n-methylpyrrolidone (NMP) at 60 °C for approximately 60 min during which the sample was gently rinsed with flowing NMP. The lift-off was finished with 15 min sonication. No damage to the nanostructures was observed after sonication. The union of multiple connected bow-ties formed a larger structure compared to the isolated nanoparticles, making the bow-ties unaffected by sonication. After lift-off, the sample was rinsed with acetone and IPA. Finally, gentle oxygen plasma ashing (50 W, 60 s) was applied to remove residual resist and solvents. The contact pads were fabricated via a subsequent photolithography step. Positive-tone photoresist S1813 (*Shipley*) was spin-coated and soft-baked at 110 °C for 4 min. Photolithography was performed with a Heidelberg µPG 101 direct laser writing system. After exposure, the samples were developed in Microposit MF-321 developer for 90 s and in DI water for 15 s. 20 nm Ti and 80 nm Au were then deposited via electron-beam evaporation. Lift-off was performed by soaking the samples in acetone for ∼30 min followed by 3 min sonication. Nanoantenna arrays consisting of 24×12 or 24×24 bow-ties were fabricated and tested. The array pitch was 800 nm in the x-direction (the direction of the bow-tie long axis), and 800 nm or 400 nm in the y-direction (the direction of the bow-tie short axis), with the full array covering an area of about 20×10 µm$^2$. For the nanotriangles, the nominal altitude was 260 nm, and the nominal base width was varied from 155 nm to 235 nm for tuning the nanoantenna plasmonic resonance. The bow-tie nano-gap size was tuned by changing the lithographic dose, with the smallest average gap size below 20 nm. The nominal linewidth for the connecting wires was 50 nm. The nominal thickness was 20 nm for both the nanoantennas and the connecting wires.

**Electromigration**

The electromigration process was similar to the one described in Ref.[39] originally used for the fabrication of metallic electrodes with nanometer separation. In our electromigration process, a bias voltage was applied to the nanoantenna array connected in series with a 2.5 Ω resistor. A small resistor ensured a small change of the voltage across the nanoantenna device during electromigration when a constant bias voltage was used. The bias voltage and the voltage (hence the current) across the resistor were monitored by an oscilloscope. The shorted devices had a low resistance and hence a high current, which broke the connecting wires of these devices via electromigration. The normal devices had a large resistance and negligible current, and remained intact after the electromigration process. As an example, Fig. 2a shows the voltage and current across a nanoantenna array during electromigration. The applied voltage (across the array and the resistor) was kept at 1 V for 50 s, and then increased to 1.25 V. Initially, the nanoantenna array was shorted, with a resistance of 15 Ω. Electromigration process started at 50 s, showing a decrease of the current, which indicates an increasing resistance. At 250 s, the current dropped to zero, implying the array was transformed into an open-circuit.

**Electromagnetic simulation**

We simulated the optical response of the plasmonic nanoantenna arrays with a finite element method electromagnetic solver (*COMSOL Multiphysics*). The modeled nanoantenna geometry was taken from layout design parameters or SEM images of fabricated nanostructures. The 20-nm-thick gold nanoantenna was placed on the interface between vacuum and a glass (BK7) substrate, with a 2-nm-thick Ti adhesion layer in between. The optical properties of Au and Ti were taken from the work by Johnson and Christy[51] describing optical constants of the metals fabricated under similar conditions to ours (vacuum-evaporated polycrystalline thin films). The refractive index of glass was fixed at 1.5 as its dispersion was negligible in the wavelength range of interest. Periodic boundary conditions were applied to the simulation domain boundaries to model a periodic array of nanoantennas. The array pitch was 800 nm in the bow-tie long-axis direction and 400 nm in the bow-tie short-axis direction. The plane-wave light was incident normally with a linear polarization along the bow-tie long-axis to excite the plasmonic mode. Perfect matched layers were added to the top and bottom of the simulation domain to absorb outgoing electromagnetic waves and model semi-infinite vacuum and substrate. Extinction and field-enhancement were evaluated in the frequency-domain. The extinction is defined as $-10log_{10}(T/T_0)$, where $T$ is the power transmissivity when the nanoantenna is present, and $T_0$ is the power transmissivity when the nanoantenna is absent (but the substrate is still present). The field-enhancement is defined as the ratio of the optical near-field at the nanotriangle tip near the nano-gap, averaged over the curved surface defined by the tip radius of curvature and gold thickness, to the optical field of the incident light. For the time-domain response, a $cos^2$-shaped incident pulse with a central wavelength of 1177 nm and a pulse duration of 10 fs FWHM was assumed. The spectrum of the pulse was obtained by a Fourier transform. Broadband (800 nm - 1600 nm wavelength) frequency-domain simulations were performed to evaluate the enhanced optical near-field at the nanotriangle tip. The field-enhancement was assumed to be unity for wavelengths outside the simulation spectral range. The time-domain response was obtained by an inverse Fourier transform of the frequency-domain response. The CEP-sensitive photocurrent was estimated by a harmonic analysis of the Fowler-Nordheim photoemission current induced by the transient optical field[19].

**Experimental setup**
The nanoantenna devices were exposed to a few-cycle, CEP-stabilized optical pulse train from a supercontinuum-based fiber laser source[41]. The supercontinuum source has a central wavelength of ≈1177 nm, pulse duration of ≈ 10 fs FWHM (≈ 2.5 cycles FWHM), repetition rate of 78 MHz, and peak pulse energy of ≈ 190 pJ. The laser beam was focused to a spot size of 2.25 μm ×4.1 μm FWHM resulting in a peak intensity of $\approx 2.6 \times 10^{11}$ W/cm$^2$ before plasmonic enhancement. In the experiments, the carrier-envelope-offset frequency $f_{ceo}$ was stabilized to 100 Hz by a local oscillator. The CEP RMS noise of the laser was about 150 mrad.

In the external circuit, the photocurrent generated by the nanoantenna device was first amplified by a transimpedance amplifier (*FEMTO*) and then detected by a lock-in amplifier (*Stanford Research Systems*) using the carrier-envelope-offset frequency as the reference frequency (see Supplementary Note 4). Discrete shifts of the CEP were introduced by the mismatch of the group and phase velocities in the BaF$_2$ wedge (2 mm thickness and 0.75° angle), which was inserted by 2.5 mm every 20 s and led to a CEP shift calculated as 57.9°. The CEP-response across entire nanoantenna arrays were measured by scanning the piezoelectric sample stage while simultaneously recording the photocurrent magnitude and phase. The photocurrent spectra were measured by a vector signal analyzer (*Agilent*). The extinction spectra of the samples were measured by a fiber-coupled optical spectrum analyzer (*Ando*). The measurements of $I_{0,collected}$ were performed by chopping the beam and using lock-in detection to measure the amplitude and phase of the first harmonic of the chopped photocurrent signal.

**Acknowledgements**

This material is based upon work supported by the Air Force Office of Scientific Research under award numbers FA9550-19-1-0065, and FA9550-18-1-0436.

# Supplementary Information:
# Light Phase Detection with On-Chip Petahertz Electronic Networks

**Yujia Yang**[1], **Marco Turchetti**[1], **Praful Vasireddy**[1], **William P. Putnam**[1,2,3], Oliver Karnbach[1], Alberto Nardi[1], **Franz X. Kärtner**[1,3,4], **Karl K. Berggren**[1], **Phillip D. Keathley**[1]

[1]Research Laboratory of Electronics, Massachusetts Institute of Technology, Cambridge, MA, USA
[2]Department of Electrical and Computer Engineering, University of California, Davis, Davis, CA, USA
[3]Department of Physics and Center for Ultrafast Imaging, University of Hamburg, Hamburg, Germany
[4]Center for Free-Electron Laser Science and Deutsches Elektronen-Synchrotron (DESY), Hamburg, Germany

**Supplementary Note 1: Determination of the bow-tie nano-gap size**

The bow-tie nano-gap sizes were measured from SEM images. To determine the gap size, we first performed a line scan across the bow-tie nano-gap (Supplementary Fig. 1a). From the grayscale value along the line, the nano-gap size was determined as the FWHM gap size (Supplementary Fig. 1b).

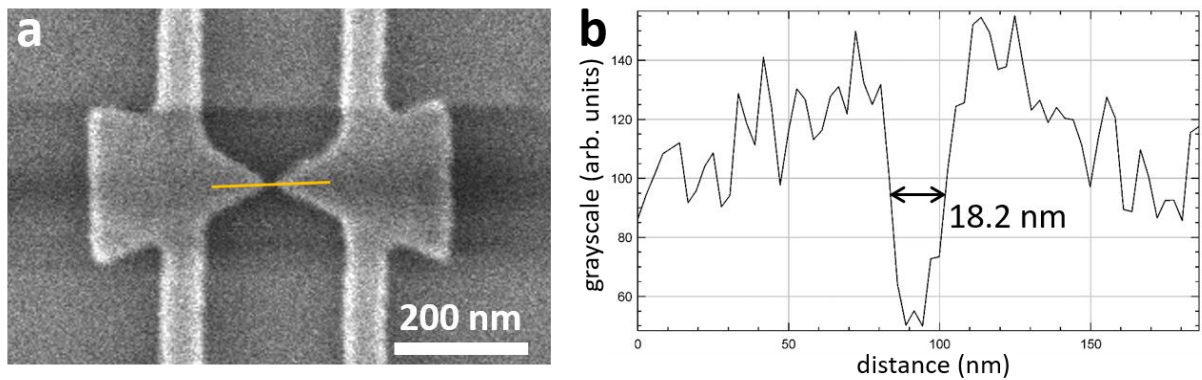

**Supplementary Figure 1**. Determination of the bow-tie nano-gap size. **a**, SEM image of a fabricated bow-tie nanoantenna. **b**, Grayscale value along a line scan (yellow line in **a**) across the nano-gap. The FWHM gap size is 18.2 nm.

**Supplementary Note 2: Further details of optical simulation of nanoantennas**

Optical simulation of the nanoantenna arrays in Fig. 1 used dimensions taken from the SEM images of a fabricated sample. The nanotriangle altitude was 245 nm, the base was 190 nm, and the bow-tie nano-gap was 50 nm. The sharp corners of the nanotriangle were rounded to avoid singularities and to better imitate fabricated samples.

Supplementary Fig. 2 shows the simulated extinction spectra, besides the field-enhancement spectra shown in Fig. 1f, of the plasmonic bow-tie nanoantenna arrays with different connecting wire positions. When the connecting wires are added, the extinction spectra (Supplementary Fig. 2) splits into two peaks, corresponding to the bow-tie and wire modes. The spectral separation of the two peaks is small and they merge into a single peak, if the connecting wire position is near the center of the nanotriangle (e.g. $X_{wire}$ = 120 nm). The spectral separation of the two peaks increases, with the bow-tie mode being blue-shifted and the wire mode being red-shifted, when the connecting wire position is close to the nanotriangle tip (e.g. $X_{wire}$ = 50 nm) or nanotriangle base (e.g. $X_{wire}$ = 200 nm). As expected from the discussions in the main text, placing the connecting wire close to the center of the nanotriangle leads to minimal perturbation of the bow-tie plasmonic mode. There is also a slight shift between the extinction peaks with respect to the field-enhancement peaks. For photoemission from nanoantenna devices, we tuned the field-enhancement peaks so that they were close to the central wavelength of the excitation laser.

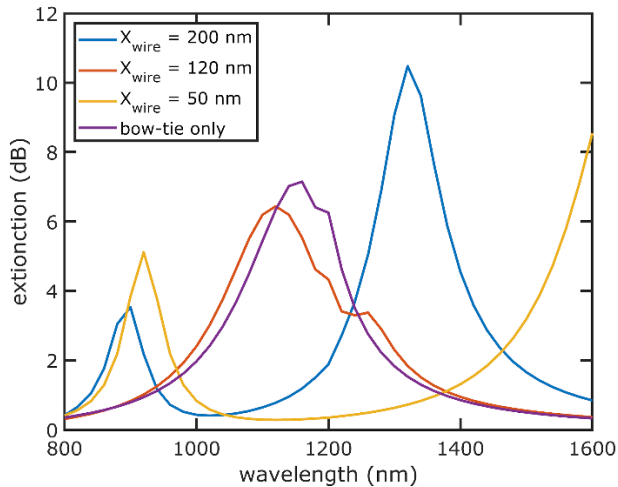

**Supplementary Figure 2**. Simulated extinction spectra of the plasmonic bow-tie nanoantenna arrays with different connecting wire positions ($X_{wire}$ labeled in Fig. 1b).

In the optical simulation with a parametric sweep of the connecting wire position (Fig. 1f&g, Supplementary Fig. 2), the thin Ti adhesion layer was neglected to reduce the computation time. It has been demonstrated that a Ti adhesion layer could cause damping of the plasmonic resonance and reduce photoemission current from metallic nanostructures[52]. However, the spectral position of the plasmonic resonance is less affected by the adhesion layer. We investigated the effect of the Ti adhesion layer with optical simulations of electrically-connected bow-tie nanoantenna arrays (Supplementary Fig. 3). In agreement with previous reports, the plasmonic resonance intensity decreases when a Ti adhesion layer is present, while the spectral position of the resonance is less affected. Therefore, optical simulation of nanoantennas without an adhesion layer is sufficient to study the effect of the connecting wire position.

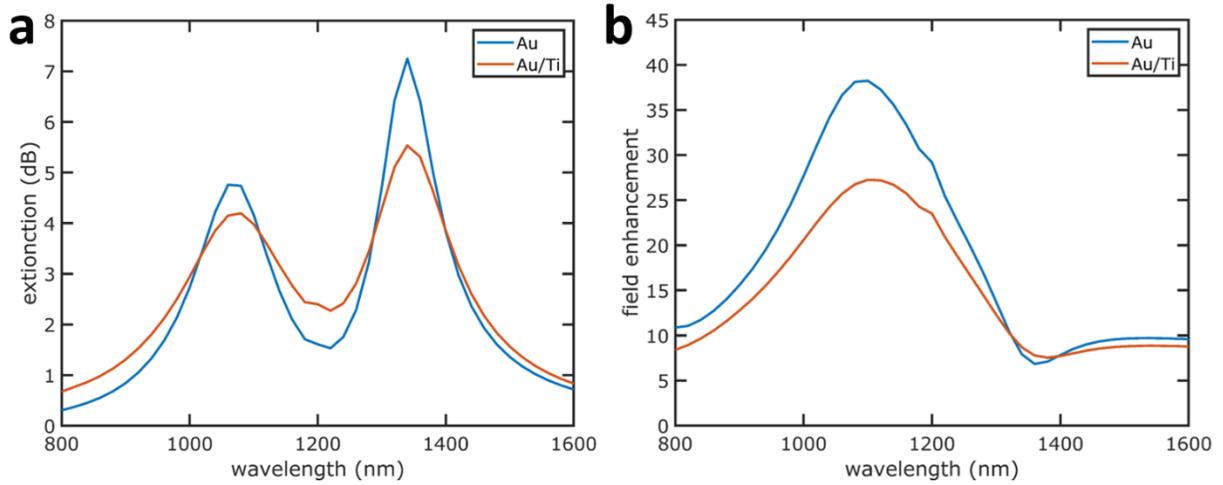

**Supplementary Figure 3**. Effect of the Ti adhesion layer on the plasmonic resonance of electrically-connected bow-tie nanoantennas. **a**, Extinction spectra, and **b**, field-enhancement spectra of the nanoantenna with (Au/Ti) and without (Au) a 2-nm-thick Ti adhesion layer.

**Supplementary Note 3: Electromigration of electrically shorted nanoantennas**

We performed electromigration on the nanoantenna arrays shorted by connected bow-ties. The electromigration results largely depended on the number of connected bow-ties within the array. For an array with most of the bow-ties connected, all columns were shorted and the electromigration broke all connecting wires, leading to a complete open-circuit array (Fig. 2b and Supplementary Fig. 4a). For an array with just a few connected bow-ties and shorted columns, electromigration removed the shorted columns and kept the open-circuit columns, enabling the CEP-sensitivity of the array to be measured (Fig. 2c&d). For the connecting wires broken by electromigration, the break usually occurred at a position close to the contact pad (Fig. 2b-d and Supplementary Fig. 4a), while occasionally the break occurred at a position between the bow-ties (Supplementary Fig. 4b).

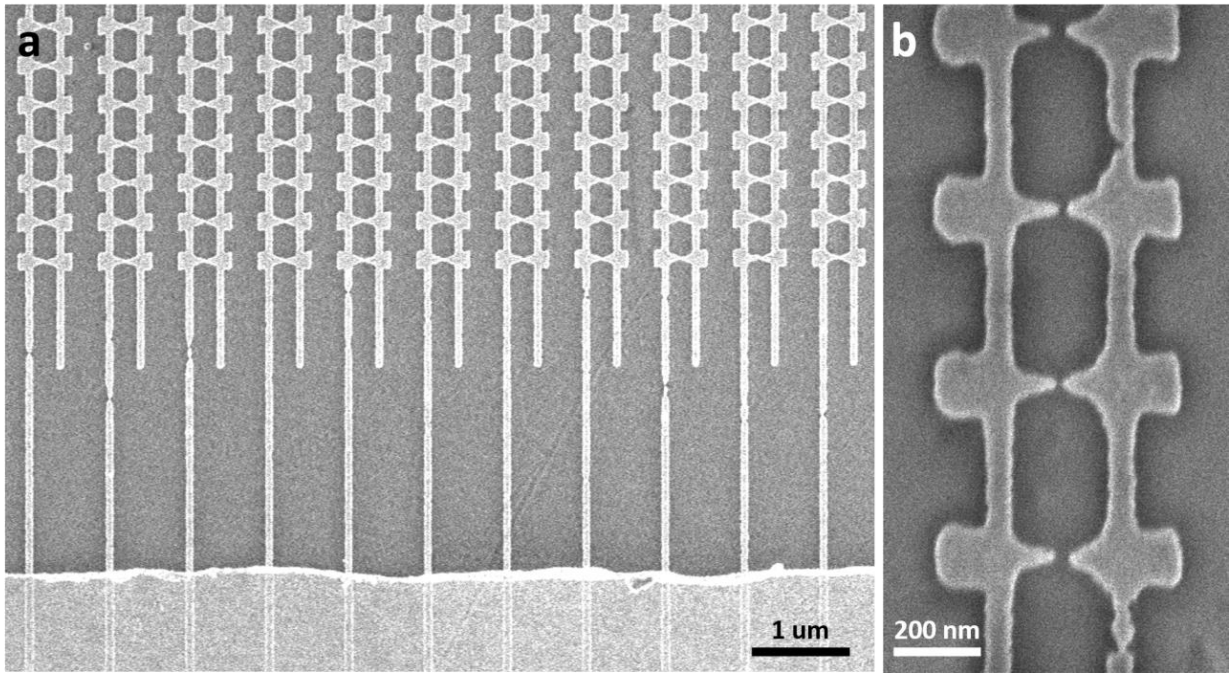

**Supplementary Figure 4**. Electromigration of electrically connected nanoantenna arrays. **a**, SEM image of a connected plasmonic nanoantenna array after electromigration. The electrical connecting wires were broken and disconnected during electromigration. **b**, SEM image of a connecting wire broken by electromigration with the breaks between the bow-tie nanostructures.

**Supplementary Note 4: Experimental setup**

Supplementary Fig. 5 shows a schematic diagram of the experimental setup.

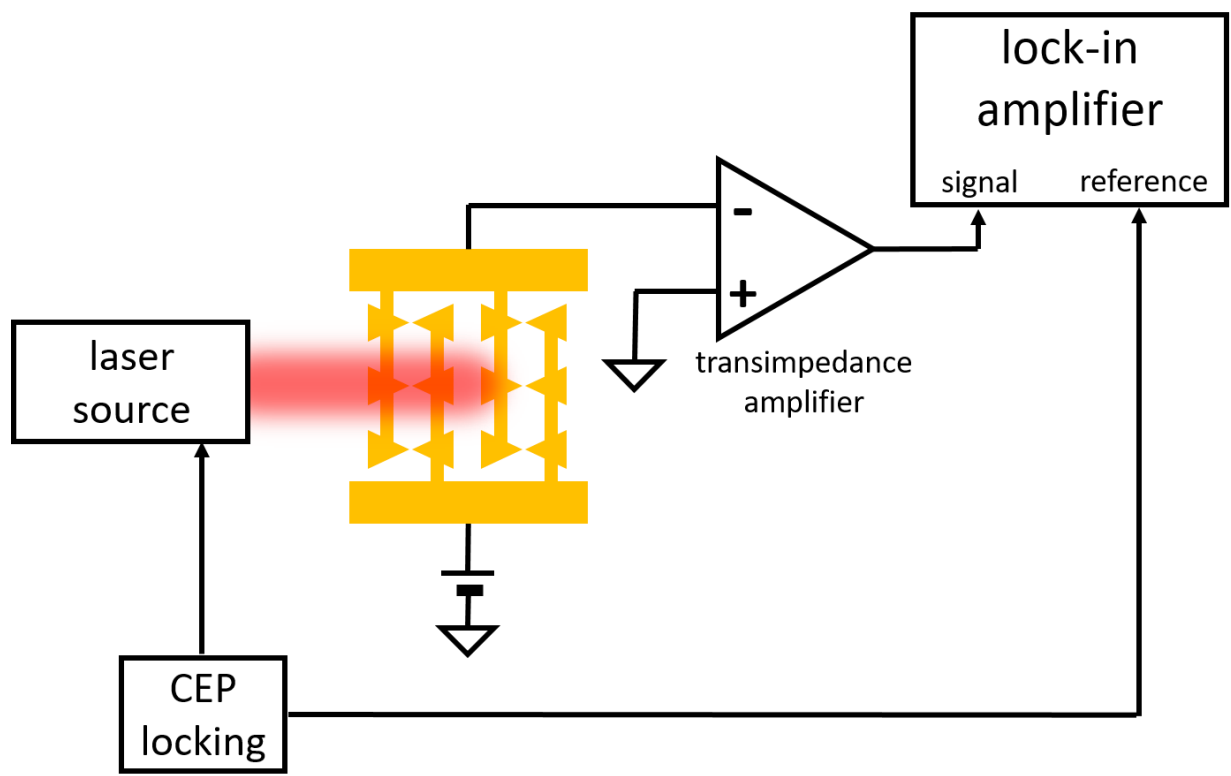

**Supplementary Figure 5.** Schematic diagram of the experimental setup. The photocurrent generated by the nanoantenna device was first amplified by a transimpedance amplifier, and then detected by a lock-in amplifier using the carrier-envelope-offset frequency as the reference frequency.

**Supplementary Note 5: Ensuring adequate intensity for optical-field emission**

To achieve CEP-sensitive emission, peak intensities are required such that the Keldysh parameter[53] $\gamma = \sqrt{\phi/2U_p} < 1$, where $\phi$ is the emitter material work function ($\approx$ 5.1eV for Au) and $U_p = e^2 F^2 \lambda^2 / 16\pi^2 c^2 m$ is the ponderomotive potential of the optical field at the emitter tip surface, where $e$ is the electron charge, $F$ the peak optical-field strength, $\lambda$ the central wavelength of the optical pulse, $c$ the speed of light, and $m$ the electron mass. Even after obtaining a field-enhancement between 20-30 × for our plasmonic bow-tie antennas, the energy limitations of our current source require tight focusing to achieve $\gamma < 1$, meaning that we are not able to characterize CEP-sensitive photoemission while illuminating the entire emitter array.

**Supplementary Note 6: Scaling CEP-sensitive signal and SNR with larger arrays and higher-energy pulses**

Assuming the peak signal measured in the experiment, we find that the given laser spot produces around 1 electron of CEP-sensitive charge per pulse, however after operating the devices for some time and scanning over the detector area, this peak signal degrades and we find an average CEP-sensitive signal of ≈ 0.1 electrons per pulse. The nanoantenna devices could operate stably at the degraded signal level for hours without further degradation. Taking into account our current laser spot size, we find that an array of size 50×50 µm$^2$ would provide a charge output of ≈ 35 electrons/pulse assuming the lower averaged emission rate as reported in the manuscript. To maintain the same peak intensity, this would require a pulse energy of ≈ 70 nJ. We find that with commercially available transimpedance amplifiers, such peak charge output could be amplified to mV-level signals sufficient for single-shot CEP tagging with an SNR improvement of more than two orders of magnitude compared to those reported in the manuscript. This would reduce the required pulse energy for single-shot CEP tagging by more than two orders of magnitude compared to ionization-based single-shot CEP detection methods. Assuming device degradation and uniformity issues could be further improved, this required pulse energy may even be further reduced by at least an order of magnitude based on the peak CEP-sensitive signals measured in this work.

Regarding the scaling of SNR of the CEP-sensitive current, we find that as long as $f_{ceo}$ is high enough to be in the shot-noise limited regime due to the total emitted current, which scales linearly with $I_{cep}$, the SNR should scale linearly with the signal, and thus the array size, and inversely with the resolution bandwidth. Using a 3 Hz resolution bandwidth, we measured SNR values ranging from 20-30 dB when referenced to the shot-noise floor (see for instance the results in Supplementary Fig. 8). A 50×50 µm$^2$ array is roughly 350× larger than the current beam spot, meaning that by illuminating an entire 50×50 µm$^2$ we could maintain a 20-30 dB SNR over a resolution bandwidth of approximately 1050 Hz.

**Supplementary Note 7: Potential avenues for further improving device SNR**

We note that for the room-temperature devices we studied in this work with gaps much larger than the deBroglie wavelength of the ground state electrons inside the nanoantennas, the emitted electron current from each triangle in the bow-tie pair was not strictly correlated, meaning that the ultimate noise floor was set by the shot-noise arising from the total emitted current from each triangle. Given the low CEP-sensitivity expected for the pulses used in this work (on the order of $10^{-4}$ to $10^{-5}$), the total emitted current is significantly larger than the CEP-sensitive current. However, if the pulse at the tip surface could be sufficiently shortened, it would be possible to significantly enhance the CEP sensitivity and thus improve the observed SNR.

Another interesting avenue might be to pursue very short gaps within superconducting antennas such that a single electron state extending across the nanoantenna gap contributes to the tunneling current. In such a case, the emission from each triangle in the bow-tie gap, and thus its resultant noise, would be correlated, thus drastically reducing the shot-noise due to the total emitted current.

**Supplementary Note 8: CEP-sensitive signals from more test devices**

We measured the CEP-sensitivity of several samples from multiple fabrication batches, and obtained similar CEP-sensitive responses. Supplementary Fig. 6a&b show the phase and magnitude of $I_{cep}$ from a nanoantenna device (Array 3) as a function of time with a laser pulse energy of 175 pJ. Discrete phase shifts were observed while translating the $BaF_2$ wedge every 20 s. The measured CEP-sensitive current magnitude had a peak value above 5 pA, and it degraded to ~2 pA. Supplementary Fig. 6c&d show the phase and magnitude of $I_{cep}$ while scanning the laser beam over Array 3. Similar to Fig. 3, the phase response of the array was relatively uniform, while there were hot and cold spots of the current magnitude. As a comparison, Supplementary Fig. 6e&f show the phase and magnitude of $I_{cep}$ from another nanoantenna device (Array 4) with a larger nano-gap size. The measured CEP-sensitive current magnitude was ~0.1 pA, and the phase response was noisier. A larger nano-gap led to a lower field-enhancement, and hence a lower CEP-sensitive photocurrent. However, the CEP-sensitive signal showed less degradation, possibly due to a reduced laser-reshaping effect resulting from a lower plasmonic enhancement.

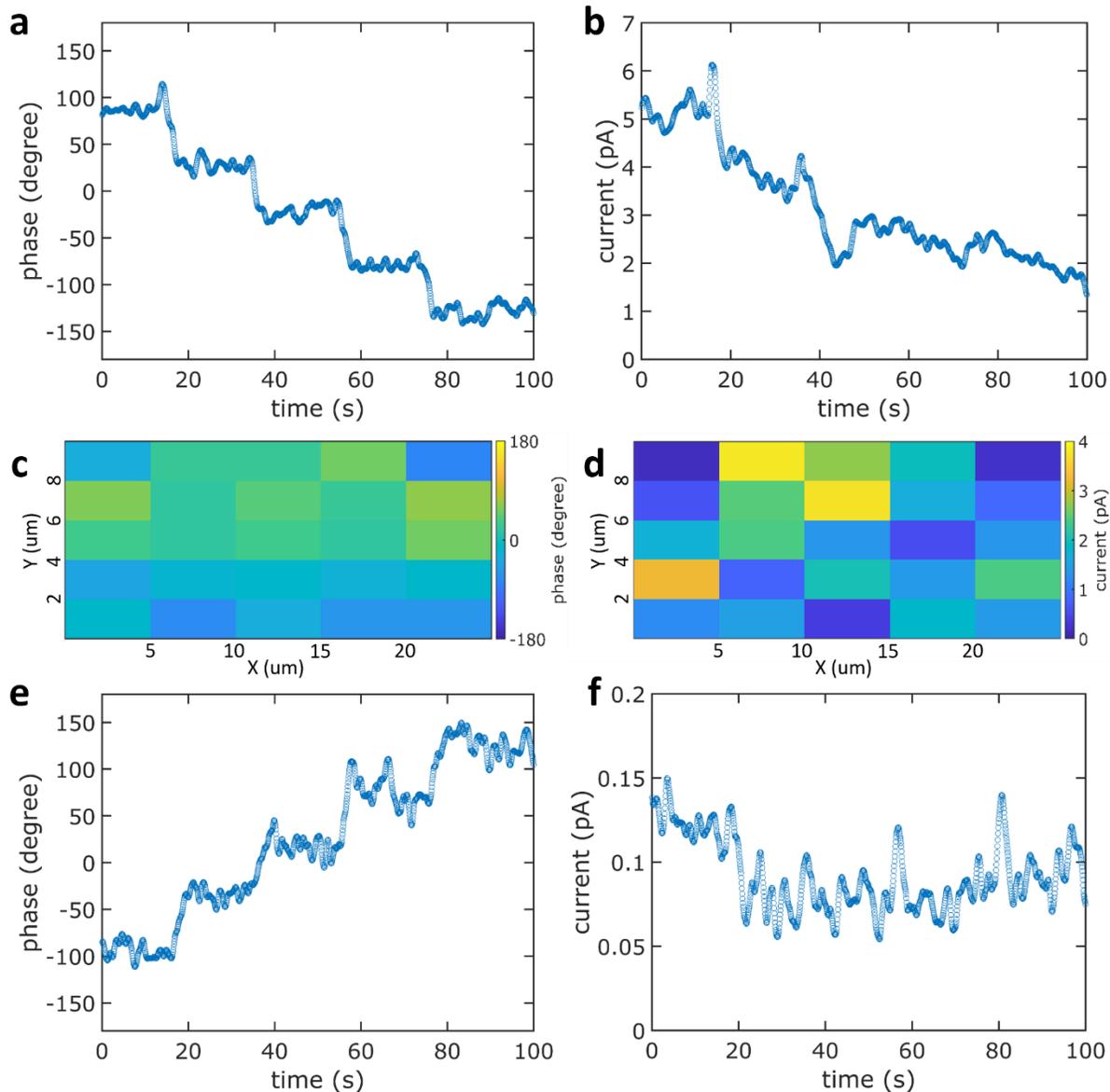

**Supplementary Figure 6**. The effect of nano-gap size on CEP-sensitive current $I_{cep}$ from nanoantenna devices. **a**&**b**, the phase and magnitude of $I_{cep}$ from a nanoantenna device (Array 3) as a function of time with a laser pulse energy of 175 pJ. **c**&**d**, the phase and magnitude of $I_{cep}$ while scanning the beam over Array 3. **e**&**f**, the phase and magnitude of $I_{cep}$ from another nanoantenna device (Array 4) as a function of time with a laser pulse energy of 155 pJ. For all measurement results, the $BaF_2$ wedge was stepwise inserted or retracted every 20 s.

Supplementary Fig. 7 shows the phase and magnitude of $I_{cep}$ from a nanoantenna device (Array 3) with various laser pulse energies. For a laser pulse energy of 175 pJ, the CEP-sensitive photocurrent had a peak value of ~5 pA and a stabilized value of ~2 pA, while for a laser pulse energy of 139 pJ or 123 pJ, the CEP-sensitive photocurrent was around 1 pA. Notably the CEP-sensitive signal showed little or no sign of degradation at a low pulse energy, which suggests the lower pulse energies didn't reshape the nanoantennas.

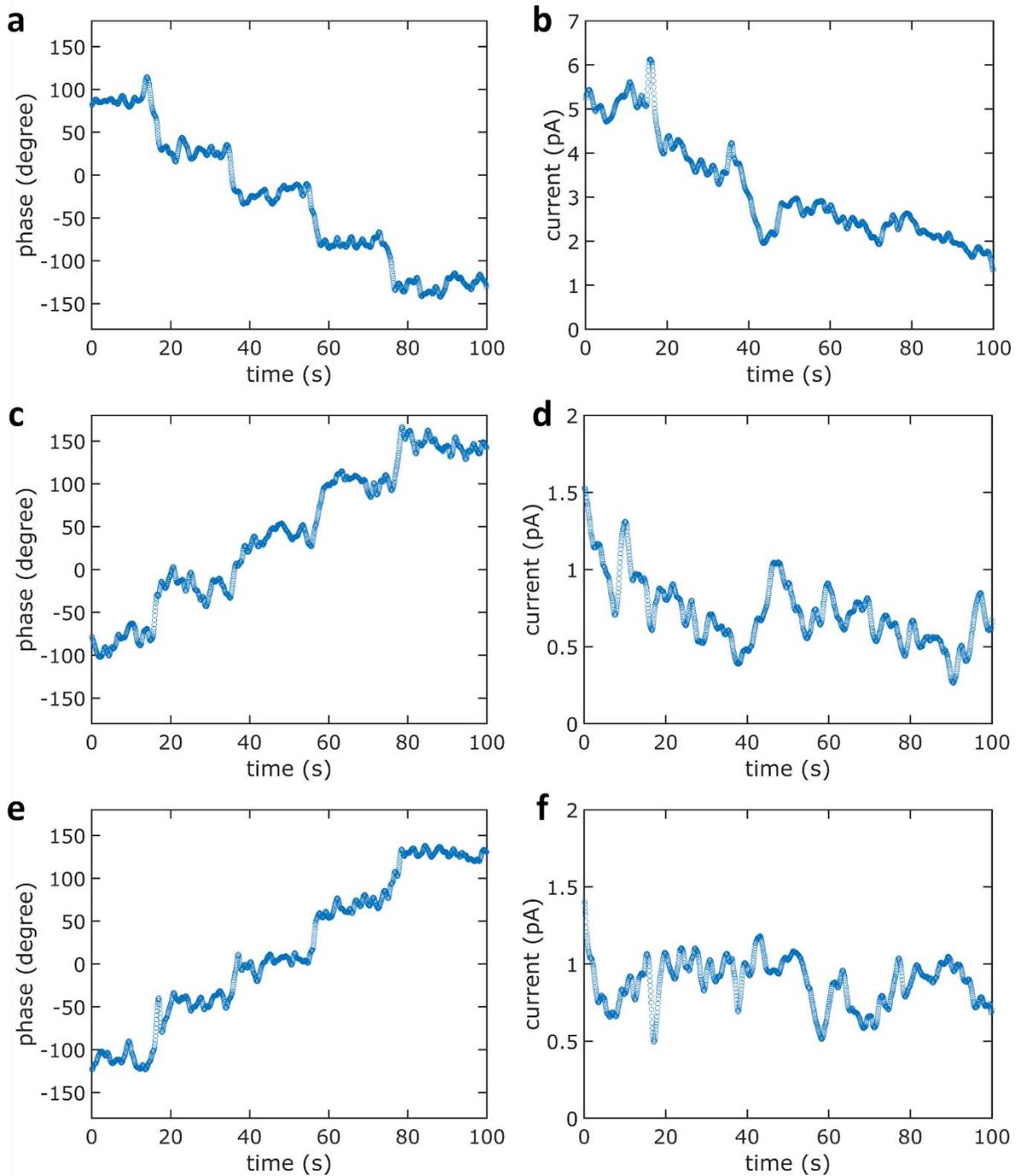

**Supplementary Figure 7**. The effect of laser pulse energy on CEP-sensitive current $I_{cep}$ from nanoantenna devices. **a**&**b**, the phase and magnitude of $I_{cep}$ from a nanoantenna device (Array 3) as a function of time with a laser pulse energy of 175 pJ (the same as Supplementary Fig. 5a&b). **c-f**, the phase and magnitude of $I_{cep}$ from the same nanoantenna device (Array 3) as a function of time with a laser pulse energy of 139 pJ (**c**&**d**) or 123 pJ (**e**&**f**). For all measurement results, the $BaF_2$ wedge was stepwise inserted or retracted every 20 s.

# Supplementary Note 9: Laser-reshaping of nanoantennas and change of CEP-sensitivity

For most of the devices tested, we observed an increase of the bow-tie nano-gap after laser illumination. Supplementary Table 1 shows the gap sizes of 3 representative samples as fabricated and after laser illumination. The laser exposure dose was roughly $10^8$ pulses with 78 MHz repetition rate and up to ~190 pJ pulse energy. Regardless of the as-fabricated gap sizes, the gap sizes after laser illumination always ended up in the 50-60 nm range. These similar gap sizes after laser illumination suggest the laser-induced reshaping was self-stabilized as the gap size increased and plasmonic enhancement decreased during laser illumination. Further investigations with more test samples and accurate laser exposure dose calibration are required to confirm this phenomenon.

| Sample # | Gap size as fabricated (nm) | Gap size after laser illumination (nm) |
|---|---|---|
| 1 | 45.0 ± 2.8 | 49.6 ± 7.1 |
| 2 | 50.5 ± 3.2 | 62.2 ± 11.9 |
| 3 | 39.3 ± 3.3 | 61.7 ± 5.7 |

**Supplementary Table 1**. Nano-gap sizes for 3 representative bow-tie nanoantenna arrays before and after laser illumination.

We observed a degraded CEP-sensitivity of the nanoantenna devices during laser illumination. This degradation could be intuitively explained as the increased bow-tie gap size leading to a decreased field-enhancement, and hence a decreased photoelectron emission current. However, it has been recently reported that the CEP-sensitivity does not monotonically change with the optical field[19]. Supplementary Fig. 8 shows the simulated CEP-sensitive photocurrent magnitude $|I_{cep}|$ with a varying peak incident optical-field strength for the nanoantenna array in Fig.4 before (blue) and after (orange) the photoemission measurement (similar to Fig. 4d but with a larger range of the incident optical field). For an incident optical field in the range of 13-40 GV/m, the nanoantenna device after illumination (with a larger nano-gap and a smaller field-enhancement) shows a higher CEP-sensitivity compared to the device before illumination, which has a vanishing CEP-sensitivity. Nevertheless, this counterintuitive behavior requires a large incident optical field (10× the optical field we used in our experiments). As we have already observed laser-induced reshaping and device degradation in the experiments, the counterintuitive, improved CEP-sensitivity after illumination is unlikely to occur in our devices.

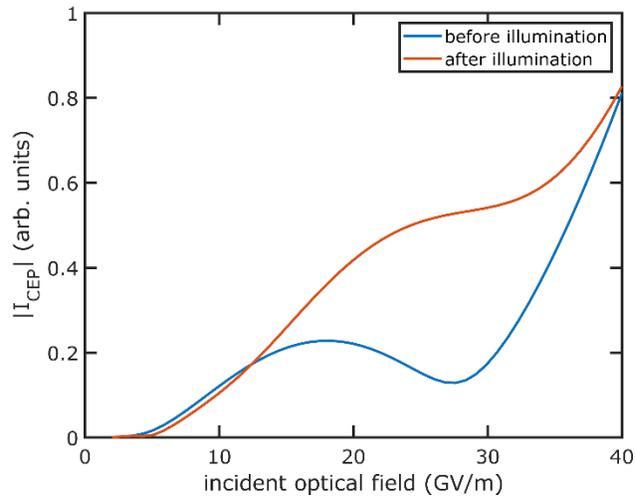

**Supplementary Figure 8**. Simulated CEP-sensitive photocurrent magnitude |$I_{cep}$| *versus* the optical near-field for the nanoantenna array before and after photoemission measurement with an incident optical field up to 40 GV/m.

**Supplementary Note 10: Photocurrent measurement with a DC bias scan**

We measured the photocurrent response at 0 Hz corresponding to the total average current collected $I_{0,\text{collected}}$ as a function of the DC bias voltage $V_{\text{bias}}$ between the two nanotriangle emitters in the bow-tie pair (e.g. Fig. 5b). Supplementary Fig. 9 shows the measurement of the total average current with a DC bias scan for two device arrays. The $V_{\text{bias}}$ value that gives $I_{0,\text{collected}} \approx 0$ A varies slightly across the samples.

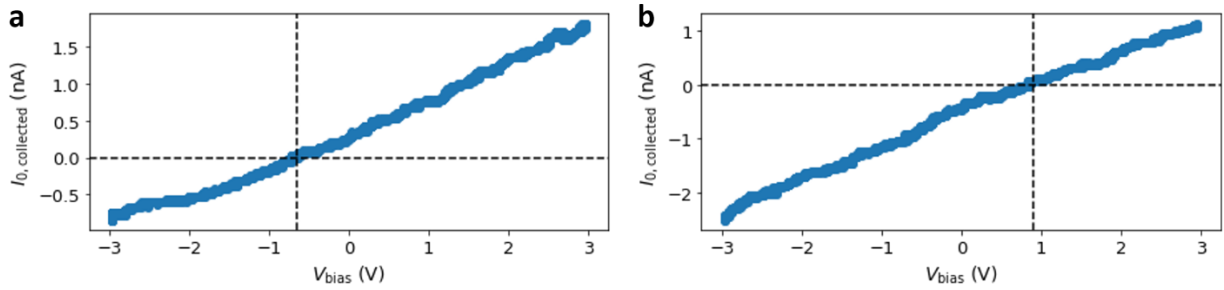

**Supplementary Figure 9**. Plots of $I_{0,\text{collected}}$ demonstrating the use of a DC bias voltage $V_{\text{bias}}$ to null the total average collected current signal. The DC bias scans for two arrays are shown (**a**, **b**). The $V_{\text{bias}}$ value that gives $I_{0,\text{collected}} \approx 0$ A varies slightly across the samples.

**Supplementary Note 11: Further investigations of the noise floor**

In the manuscript we argue that the noise arises from shot-noise resulting from $I_{0,\text{emitted}}$. To provide further evidence of this, we recorded current spectra showing the CEP beat note (here at 120 Hz) and surrounding noise floor when exposed to ≈190 pJ pulses with various bias conditions as shown in Supplementary Fig. 10. Note that the biased case is for $V_{\text{bias}}$ = 3 V, the unbiased case for $V_{\text{bias}}$ = 0 V, the unlocked case is for no bias and no CEP locking, and the noise floor was recorded with no optical beam directed onto the devices. We note that the resolution bandwidth for these measurements was 3 Hz. The noise floor was found to be on the order of just 30-40 fA, agreeing with measurements of the noise-floor using the lock-in amplifier, and around one order of magnitude reduced from the noise floor when illuminated. This confirms that our observed noise indeed arises from the devices under illumination.

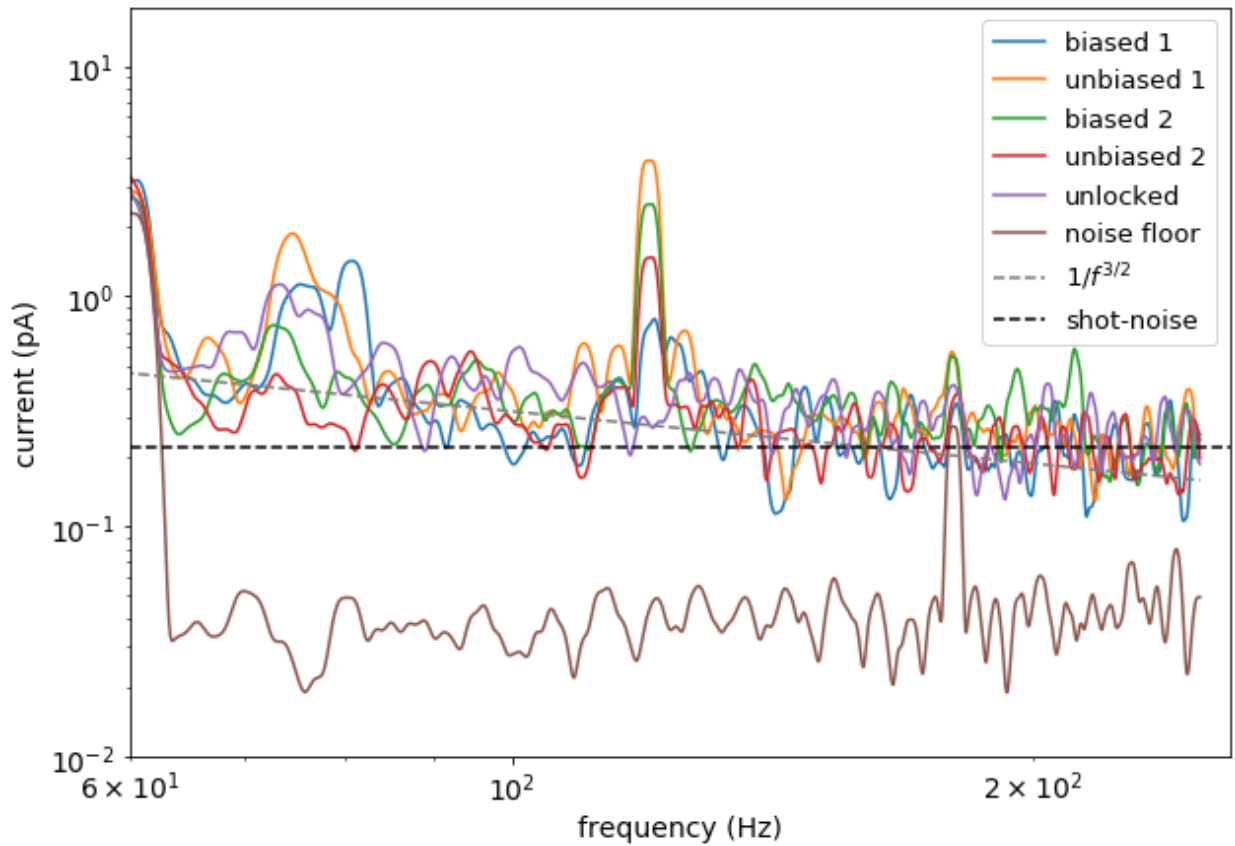

**Supplementary Figure 10**. Current spectra showing the CEP beat note (at 120 Hz here) and the background noise floor for various settings of $V_{\text{bias}}$. The biased curves are for $V_{\text{bias}}$ = 3 V and the unbiased curves for $V_{\text{bias}}$ = 0 V. Two sets of measurements were performed (biased & unbiased 1, biased & unbiased 2), and the device degradation caused a reduction of the signal peak. The unlocked case is unbiased and was taken with the CEP unlocked. The noise floor was taken with the beam blocked (*i.e.* no illumination of the devices by the laser beam). Reference curves for $1/f^{3/2}$ (dashed gray) and shot-noise (dashed black) scaling are shown indicating a transition near 150 Hz for this particular sample.

As usual, there was a slow degradation in CEP, but we found that there was no correlation between the strength of the CEP note and the bias in general up to $V_{bias}$ = 3 V. Importantly, the measurements clearly show no correlation between the noise floor and the bias despite an observed increase in $I_{0,collected}$ by several orders of magnitude when the bias is on as opposed to when it is off (see Fig. 5). This indicates that the noise arises either from the illumination or the total emitted current $I_{0,emitted}$, which would be unaltered by a mild bias.

Furthermore, we can compare the CEP response and relative noise floor to the single-triangle devices reported in Ref.[15]. In Supplementary Fig. 11, we compare single-sided triangular devices to the electrically-connected devices used in this work. The resonance of the single-sided triangles was similar to the electrically-connected devices (near 1158 nm). We purposefully chose a region of electrically-connected emitters having a similar CEP response under full illumination to the single-sided triangle devices. As shown in Supplementary Fig. 11, the SNR and noise floor are almost identical. This highlights that the dominant noise source is not common-mode in origin (for example noise due to energy fluctuations of the optical source) as common-mode noise sources should have been reduced significantly using the symmetric electrically connected devices.

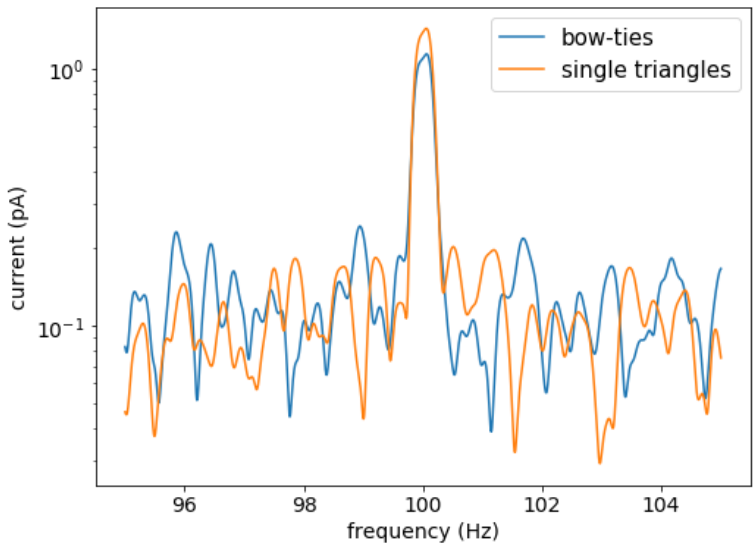

**Supplementary Figure 11.** Comparison of single triangle emitters (see Ref.[15]) to the electrically connected bow-tie devices. A region of devices was chosen such that the CEP response is similar between the two cases. They were both illuminated with the same optical conditions (peak energy near 200 pJ). Both devices exhibit nearly identical noise floor under similar conditions, emphasizing that the dominant noise source is not common-mode in origin. Note, a resolution bandwidth of 0.3 Hz was used.

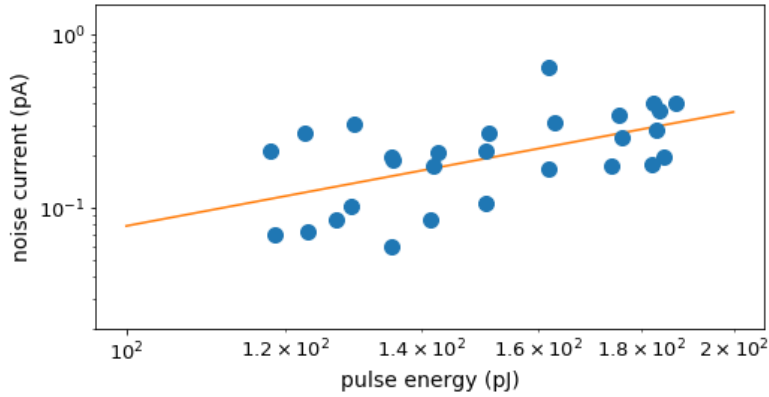

**Supplementary Figure 12**. Plot of the noise current as a function of pulse energy incident on the structures. The trend-line fit indicates a scaling proportional to approximately $P^{2.18}$. However, for intensity-induced thermal noise, one would expect a scaling with $P^{0.5}$.

Finally, the noise-floor level is examined as a function of the incident pulse energy in Supplementary Fig. 12. The data shown is in fact the same as that from Array 1 in Fig. 5 of the manuscript, only now plotted as a function of pulse energy. There are two key reasons we rule out thermal noise from absorption of the incident laser power as the primary contributor to the noise floor. First, there is visually less correlation between $P$ (the incident pulse energy) and $I_{noise}$ as between $I_{0,collected}$ and $I_{noise}$. Second, when fitting the scaling factor, we find that the noise grows proportionally to $P^{2.18}$ while one would expect laser-induced thermal noise to grow as the square-root of the temperature, and thus as $P^{0.5}$ [43].